\documentclass[prl,twocolumn,showpacs,amsmath,amssymb,superscriptaddress]{revtex4-2}

\usepackage{amsmath,amssymb,amsfonts,bm,color,graphicx,tabularx}

\usepackage[unicode=true,colorlinks=true]{hyperref}

\usepackage[etex=true,export]{adjustbox}
\usepackage{makecell}
\usepackage{multirow}
\usepackage{verbatim}

\hypersetup{linkcolor=blue, citecolor=blue,urlcolor=blue}
\usepackage{tabularx,multirow,array,diagbox}

\newcommand{\beq}{\begin{equation}}
\newcommand{\eeq}{\end{equation}}
\newcommand{\beqn}{\begin{eqnarray}}
\newcommand{\eeqn}{\end{eqnarray}}

\begin{document}

\title{Symmetry Condition and Mechanism for the Anomalous Hall Effect\\ in Odd-Parity Magnets}

\author{Xun-Jiang Luo}
\email{xjluo@hmfl.ac.cn}
\affiliation{Anhui Province Key Laboratory of Low-Energy Quantum Materials and Devices, High Magnetic Field Laboratory, HFIPS, Chinese Academy of Sciences, Hefei, Anhui 230031, China}
\author{Su-Qi Zhang}
\affiliation{College of Electrical Engineering, Naval University of Engineering, Wuhan 430033, China}
\author{Yi-Dong Liu}
\affiliation{Key Laboratory of Materials Physics, Institute of Solid State Physics,
HFIPS, Chinese Academy of Sciences, Hefei 230031, China}
\author{Yong-Ting Shi}
\affiliation{Institute of Applied Physics and Computational Mathematics, Beijing 100088, China}
\affiliation{Anhui Province Key Laboratory of Low-Energy Quantum Materials and Devices, High Magnetic Field Laboratory, HFIPS, Chinese Academy of Sciences, Hefei, Anhui 230031, China}
\author{Ding-Fu Shao}
\email{dfshao@issp.ac.cn} 
\affiliation{Key Laboratory of Materials Physics, Institute of Solid State Physics,
HFIPS, Chinese Academy of Sciences, Hefei 230031, China}
\author{Mingliang Tian}
\email{tianml@hmfl.ac.cn}
\affiliation{Anhui Province Key Laboratory of Low-Energy Quantum Materials and Devices, High Magnetic Field Laboratory, HFIPS, Chinese Academy of Sciences, Hefei, Anhui 230031, China}

\begin{abstract}
The anomalous Hall effect (AHE) in unconventional magnets has attracted
considerable attention.
A giant AHE has recently been observed in the candidate $p$-wave magnet
Gd$_3$(Ru$_{1-\delta}$Rh$_{\delta}$)$_4$Al$_{12}$ [\textcolor{blue}{Nature \textbf{646}, 837
(2025)}], while its theoretical interpretation invokes a spontaneous magnetization to
break time-reversal symmetry ($\mathcal{T}$), although the magnetic
order itself already breaks $\mathcal{T}$. Consequently,
the microscopic origin of the AHE in odd-parity magnets (OPMs)
remains an open question. Here we establish the symmetry condition and
microscopic mechanism for the AHE in OPMs. Since the OPM order is classified by the spin group whereas the AHE is governed
by the magnetic group, the AHE is allowed only when the magnetic point group is
ferromagnetic; conversely, it is forbidden whenever the effective time reversal
$\mathcal{T}\bm{\tau}$ (with $\bm{\tau}$ a fractional translation), which is essential to
the realization of the OPM, is preserved. Microscopically, OPMs generically
host a hidden Zeeman field (HZF) rooted in this intrinsic $\mathcal{T}$
breaking, and its interplay with spin-orbit coupling (SOC) generates the AHE
without any external field. We demonstrate this picture with models on the
square, honeycomb, and kagome lattices. In particular, a square-lattice
helimagnet with sublattice-resolved SOC maps exactly onto an anisotropic
Rashba--Zeeman model, whose Zeeman term is nothing but the HZF and directly
yields the AHE. Our results establish the symmetry condition for the AHE in
OPMs and identify the HZF--SOC interplay as a general route to realize it.
\end{abstract}

\maketitle

	Unconventional magnetism has emerged as a frontier of condensed matter
	physics, attracting broad attention \cite{Hayami2019,Yuan2020a,Libor2020,Ma2021,Feng2022,Tomas2022,shaodingfu2023,Krempaský2024,ZhangRunWu2024,ZhouXiaodong2024,Reimers2024,BaiLing2024,McClarty2024,LeeSuyoung2024,PhysRevLett.134.026001,HuMengli2025,Chen2025,Zhang2025a,2026arXiv260421397Y,luo2026unifiedsymmetry}. Unconventional magnets fall into two
	major classes \cite{Liu2025a,xiaobingchen2026}. The first class comprises antiferromagnets with
	nonrelativistic spin splitting (NSS), typified by the widely studied
	altermagnets with collinear magnetic order \cite{ifmmode2022,ifmmode2022a,PhysRevLett.132.056701,PhysRevLett.131.256703,PhysRevLett.132.176702,PhysRevX.14.011019,PhysRevLett.134.106801}. This class is governed entirely
	by spin-space-group symmetry, which enforces zero net magnetization but
	does not enforce spin degeneracy~\cite{LiuPengfei2022, CheXiaobing2024, JiangYi2024, XiaoZhenyu2024,zy7s-j86r,luoxun2026}. The
	second class comprises antiferromagnets that exhibit the anomalous Hall
	effect (AHE) \cite{PhysRevLett.112.017205,Ajaya2016,PhysRevB.95.075128,PhysRevMaterials.3.044409,Smejkal2022}. Its diagnosis requires
	a joint analysis of the spin space group and the magnetic point group: the
	former enforces zero net magnetization, whereas the latter must be a ferromagnetic point group for realizing the AHE \cite{Yuntian2025}. Because the two classes rest on different
	symmetry frameworks, they can intersect: a single antiferromagnet may
	simultaneously host NSS and an AHE, both of which are important for
	antiferromagnetic spintronics applications \cite{RevModPhys.90.015005,Smejkal2018,Shao2021,Shao2024,Rimmler2025}.

	Recently, odd-parity magnets (OPMs), which host an NSS that is odd under
	momentum inversion, have emerged as a new class of unconventional
	magnetism~\cite{HayamiSatoru2020,BirkHellenes2023,Brekke2024,zk69-k6b2,85fd-dmy8,YuPing2025,2025ziting,Sukhachov2025,Huang2025,2025li,2025zhu,zhuang2025,2025liu,2025luospin,2026arXiv260307643L,2026arXiv260623806L,liu2026completehierar,zhang2026oddparitymagnons,sun2026}. The simplest
	example is the $p$-wave magnet, typically realized by coplanar
	helimagnetic orders~\cite{Brekke2024}, with candidate
	materials ranging from the metal CeNiAsO \cite{2026arXiv260602422Z,zhang2026oddspin,zhang2026quench} to the insulator NiI$_2$~\cite{Song2025}. Very recently, a giant AHE was
	observed in the candidate $p$-wave magnet Gd$_3$(Ru$_{1-\delta}$Rh$_{\delta}$)$_4$Al$_{12}$~\cite{Yamada2025a}.
	The proposed interpretation \cite{Yamada2025a,okumura2026}, however, models the system as a
	Rashba-plus-Zeeman system, in which the Zeeman field stems from a
	spontaneous net magnetization that breaks $\mathcal{T}$. This picture is
	internally inconsistent: the magnetic order of an OPM already breaks
	$\mathcal{T}$ intrinsically, and this breaking manifests as a hidden
	Zeeman field (HZF)~\cite{j9m4-386b}.  This contradiction shows that the origin of the AHE
	in OPMs remains an open question.

	In this Letter, we establish the symmetry condition and the microscopic
	mechanism of the AHE in OPMs. Our central observation is that an OPM is
	defined by the spin group of the system without SOC \cite{2025luospin}, whereas the AHE is
	decided by the magnetic group with SOC. The AHE in OPMs can therefore be
	realized when the spin-group symmetry criteria for OPMs are satisfied and
	the magnetic point group of the system is ferromagnetic. This requirement
	excludes the effective time-reversal symmetry $\mathcal{T}\bm{\tau}$ (with
	$\bm{\tau}$ a fractional translation) and implies that the AHE in OPMs can be
	realized only when the system has a coplanar magnetic order without
	$\mathcal{T}\bm{\tau}$ symmetry. When the magnetic point group is
	ferromagnetic, SOC breaks the residual spin-group symmetry
	$\{\mathcal{T}U_z(\pi)\,|\,I\}$ of the coplanar magnetic order (with $U_z(\pi)$ a $\pi$ spin rotation about the $z$ axis and $I$ the identity), and its
	interplay with the HZF~\cite{j9m4-386b}, rooted in the intrinsic
	$\mathcal{T}$ breaking, yields the AHE. We substantiate this picture
	analytically for a two-dimensional (2D) coplanar helimagnet with
	sublattice-resolved SOC, which can be mapped exactly onto an anisotropic
	two-band Rashba--Zeeman model, and numerically for OPMs on the
	honeycomb and kagome lattices,
	which exhibit an $f$-wave spin texture. Furthermore, we identify candidate
	OPM materials that could exhibit the AHE. Our work establishes OPMs as a
	symmetry-guided platform for anomalous Hall physics.

\begin{figure*}
\centering
\includegraphics[width=6.9in]{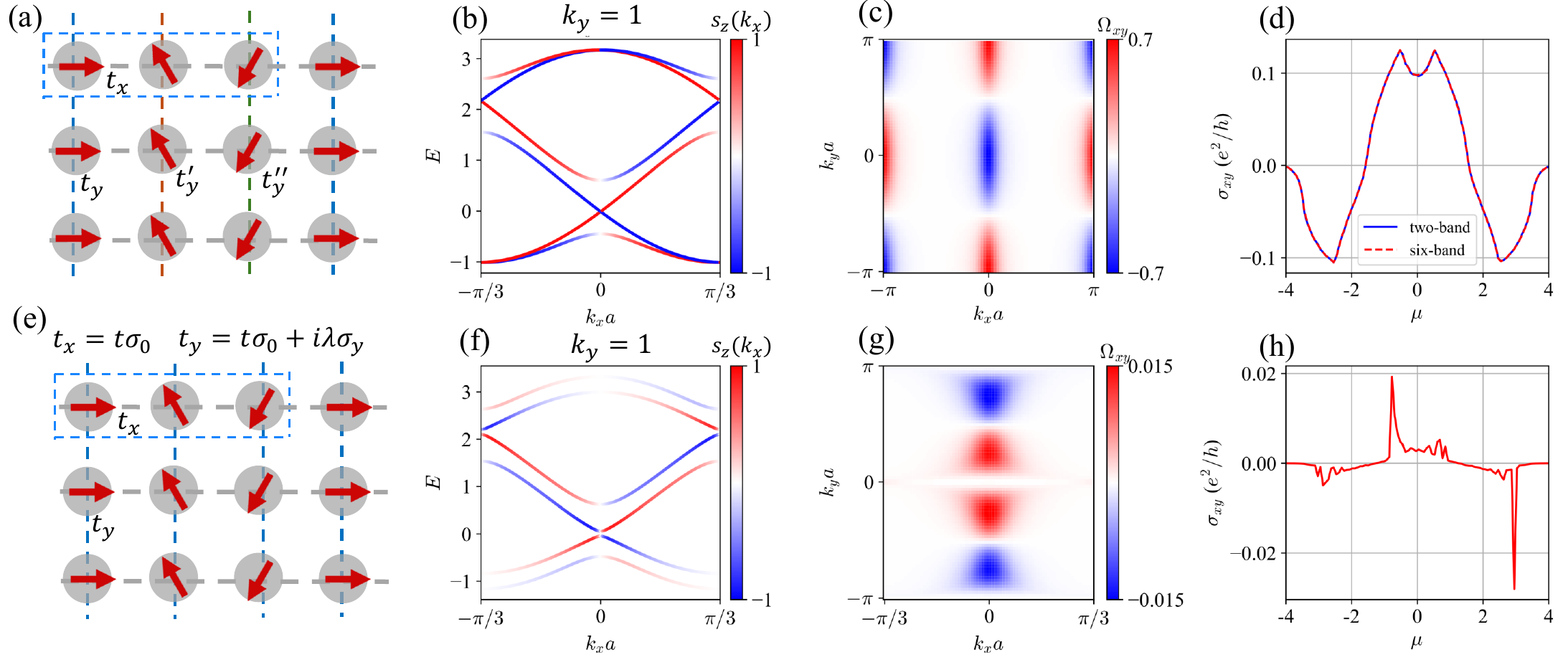}
\caption{AHE in the 2D helimagnet. (a) Schematic of the model: 1D $120^\circ$ helical chains along $x$ coupled along $y$, with sublattice-resolved interchain SOC hoppings $t_y$, $t_y'$, $t_y''$ [Eq.~(\ref{eq:SOC_momentum})]. (b) Six-band spectrum at $k_ya=1$, colored by the spin polarization $s_z(k_x)$ ($t=1$, $J=0.5$, $\lambda=0.1$). (c) Momentum-space distribution of the Berry curvature $\Omega_{xy}(\bm k)$ of the occupied bands. (d) Anomalous Hall conductivity $\sigma_{xy}(\mu)$ of the full six-band model (dashed red) and of the unfolded two-band model $\mathcal{H}(\bm k)$ of Eq.~(\ref{eq4}) (solid blue). (e)--(h) Same as (a)--(d) but for a uniform bond SOC $t_y=t\sigma_0+i\lambda\sigma_y$. The AHE survives with a reduced magnitude, demonstrating that the mechanism is independent of the specific SOC form.}
\label{Fig1}
\end{figure*}

\textit{Symmetry condition for the AHE in OPMs.}---The spin-group symmetry criteria for OPMs were established in Ref.~\cite{2025luospin}, where OPMs were classified into three types: type-I, type-II, and type-III OPMs host collinear, coplanar, and noncoplanar spin textures, respectively. Whether the AHE can be realized in these three types of OPMs, however, depends entirely on the magnetic point group of the system: the AHE is allowed when the magnetic point group is ferromagnetic and forbidden otherwise. Therefore, the symmetry $\mathcal{T}\bm{\tau}$, which places the system in a gray magnetic group, must be broken to realize the AHE in OPMs. Notably, type-II and type-III OPMs generally preserve $\mathcal{T}\bm{\tau}$ and therefore cannot host an AHE \cite{2025luospin}. Type-I OPMs are typically realized by a coplanar magnetic order (taken to lie in the $x$--$y$ plane) when the system preserves the spin-group symmetry $\{U_z(\theta)||I|\bm{\tau}\}$. When $\theta=\pi$, the system additionally preserves $\mathcal{T}\bm{\tau}$, which can be viewed as the composition of $\{U_z(\theta)||I|\bm{\tau}\}$ and the spin-only symmetry $\{\mathcal{T}U_z(\pi)\,|\,I\}$, and the AHE is again forbidden. Consequently, the AHE in OPMs can be realized only in coplanar magnetic orders without $\mathcal{T}\bm{\tau}$ symmetry, a condition that excludes most candidate OPM materials. In the Appendix A, we list the candidate OPMs that could host the AHE according to the established symmetry condition.

\textit{Effective Rashba--Zeeman model from a 2D helimagnet.}---To establish the microscopic mechanism for the AHE in OPMs, we analytically study a 2D helimagnet (a $p$-wave magnet) formed by stacking 1D helical chains along $x$ and coupling them along $y$ [Fig.~\ref{Fig1}(a)]. The magnetic unit cell contains three sublattices A, B, and C, with the magnetic moments rotating by \(120^\circ\) between neighboring sites along the $x$ direction. 
The Bloch Hamiltonian in the combined sublattice--spin basis reads
\begin{equation}
H(\bm k) = H_{\rm ch}(k_x) + 2t\cos k_y \, I_{6} + H_{\rm SOC}(k_y),
\label{eq:H_2D_total}
\end{equation}
where $H_{\rm ch}(k_x)$ is the $6\times6$ Bloch Hamiltonian of a single chain and reads
\begin{equation}
H_1(k_x) = \begin{bmatrix}
h_{\text{AA}} & T & T^{\dagger} \\
T^{\dagger} & h_{\text{BB}} & T \\
T & T^{\dagger} & h_{\text{CC}}
\end{bmatrix},
\label{eq:original_H}
\end{equation}
Here, \(h_{\text{AA}}=J \sigma_x\), \(h_{\text{BB}}=J(-\sigma_x/2 + \sqrt{3}\sigma_y/2)\), \(h_{\text{CC}}=-J(\sigma_x/2 + \sqrt{3}\sigma_y/2)\), and \(T=t e^{ik_x}\sigma_0\) with \(k_xa\in[-\pi/3,\pi/3]\), where the Pauli matrices act on the spin space and $\sigma_0$ is the $2\times2$ identity matrix; $a$ is the lattice constant and $J$ denotes the strength of the exchange coupling. The second term in Eq.~\eqref{eq:H_2D_total} is the interchain hopping, and we introduce a sublattice-resolved SOC along $y$,
\begin{equation}
H_{\rm SOC}(k_y) = -2\lambda \sin k_y
\begin{bmatrix}
M_{\rm A} & 0 & 0 \\
0 & M_{\rm B} & 0 \\
0 & 0 & M_{\rm C}
\end{bmatrix},
\label{eq:SOC_momentum}
\end{equation}
with $M_{\rm A} = \sigma_y$, $M_{\rm B} = -\frac{1}{2}\sigma_y + \frac{\sqrt{3}}{2}\sigma_x$, and $M_{\rm C} = -\frac{1}{2}\sigma_y - \frac{\sqrt{3}}{2}\sigma_x$. $\lambda$ denotes the strength of the SOC. When $\lambda=0$, $H(\bm k)$ exhibits spin splitting, and its out-of-plane spin expectation value satisfies $s_z(k_x,k_y)=-s_z(-k_x,k_y)$, so that the system behaves as a $p$-wave magnet.

We note that $H(\bm k)$ belongs to the ferromagnetic point group $1$ and therefore can support the AHE. To verify this, we show that the Hamiltonian $H(\bm k)$ can be mapped exactly onto a two-dimensional Rashba--Zeeman model by a sequence of three \emph{momentum-independent} unitary transformations, 
\beqn
&&\tilde{H}(\bm k)=UH(\bm k)U^{\dagger},\quad U=U_1U_2U_3  \nonumber\\
&&\tilde{H}(\bm {\tilde{k}})=\epsilon_0(\bm {\tilde{k}})\sigma_0+\bm d(\bm {\tilde{k}})\cdot\tilde{\bm\sigma},
\label{eq4}
\eeqn 
where $\bm k=(\tilde{k}_x,k_y)$ in the original Brillouin zone of the square lattice, $\tilde{\bm \sigma}=U\bm \sigma U^{\dagger}$, and
\begin{align}
\epsilon_0(\bm k) &= 2t \cos k_y - t \cos \tilde{k}_x,  \nonumber\\
\bm{d}(\bm{k}) &= \bigl( J,\; 2\lambda \sin k_y,\; \sqrt{3}\,t \sin \tilde{k}_x \bigr).
\label{eq5}
\end{align}
The first transformation \(U_1\) aligns the moments on all sites and reads
\begin{equation}
U_1 = \operatorname{diag}\bigl(\sigma_0,\; U(\theta),\; U(2\theta)\bigr),
\label{eq:U1}
\end{equation}
where \(U(\theta) = \exp(i\theta \sigma_z)\) with $\theta=\pi/3$. Applying \(U_1\) to the original Hamiltonian yields
\begin{align}
{H}_{\text{ch}}^{\prime}(k_x) &= U_1 H_1(k_x) U_1^\dagger \nonumber \\
&= \begin{bmatrix}
J \tilde{\sigma}_x & T_1 & T_2  \\
T_1^\dagger & J \tilde{\sigma}_x & T_1 \\
T_2^\dagger & T_1^\dagger & J \tilde{\sigma}_x
\end{bmatrix},
\label{eq:H1}
\end{align}
where \(T_1 = t e^{ik_x} e^{-i\theta \tilde{\sigma}_z}\) and \(T_2 = t e^{-ik_x} e^{-2i\theta \tilde{\sigma}_z}\), and $\tilde{H}_{\rm SOC}=U_1 H_{\rm SOC}(k_y) U_1^\dagger = -2\lambda\sin k_y\, \mathrm{diag}(\tilde\sigma_y,\tilde\sigma_y,\tilde\sigma_y)$. After the transformation $U_1$, the HZF in $H_{\text{ch}}(k_x)$ becomes transparent and takes the form $J\tilde{\sigma}_x$; it reflects the intrinsic $\mathcal{T}$ breaking of the magnetic order. The unitary transformation $U_2$ is defined by 
\begin{equation}
U_2 = \operatorname{diag}(\sigma_0,\; -\sigma_0,\; \sigma_0),
\label{eq:U2}
\end{equation}
which leads to
\begin{equation}
H_1^{\prime\prime}(k_x) = U_2 H_1^{\prime}(k_x) U_2^\dagger =
\begin{bmatrix}
J \tilde{\sigma}_x & T_3 & T_3^{\dagger} \\
T_3^{\dagger} & J \tilde{\sigma}_x & T_3 \\
T_3 & T_3^{\dagger} & J \tilde{\sigma}_x
\end{bmatrix},
\label{eq:H2}
\end{equation}
with \(T_3 = -T_1 = t e^{i(k_x+\pi)} e^{-i\theta\sigma_z}=te^{ik_x}e^{2i\theta\tilde{\sigma}_z}=T_2^{\dagger}\). Now all three sublattices are connected cyclically by the same hopping operator \(T_3\); translational symmetry is restored. Notice that \(H_1^{\prime\prime}\) describes a 1D system with complex hopping $-t e^{-i\pi \tilde{\sigma}_z/3}$, which encodes the emergent gauge field of the magnetic order and accounts for the NSS of $H_{\text{ch}}(k_x)$ \cite{j9m4-386b}.

To block-diagonalize $H_1^{\prime\prime}$, we introduce the unitary transformation $U_3$
\begin{equation}
U_3 = \frac{1}{\sqrt{3}}
\begin{pmatrix}
\sigma_0 & \sigma_0 & \sigma_0 \\
\sigma_0 & \omega \sigma_0 & \omega^2 \sigma_0 \\
\sigma_0 & \omega^2 \sigma_0 & \omega \sigma_0
\end{pmatrix},
\label{eq:U3}
\end{equation}
with \(\omega = e^{-2\pi i/3}\). After this transformation, the Hamiltonian becomes block‑diagonal:
\begin{equation}
\tilde{H}_{\text{ch}}(k_x) = U_3 H_{\text{ch}}^{\prime\prime}(k_x) U_3^\dagger = \operatorname{diag}\bigl({h}_0,\; {h}_1,\; {h}_2\bigr),
\label{eq:block_diag}
\end{equation}
where each block is a \(2\times2\) matrix:
\begin{align}
\tilde{h}_m(k_x) = J \tilde{\sigma}_x + t\Bigl(e^{ik_x} \omega^m e^{-i\theta \tilde{\sigma}_z} + e^{-ik_x} \omega^{-m} e^{-i\theta \tilde{\sigma}_z}\Bigr).
\end{align}
The three blocks are not independent; they satisfy
\begin{align}
\tilde{h}_1(k_x) &= \tilde{h}_0\!\left(k_x + \frac{2\pi}{3}\right), \nonumber\\
\tilde{h}_2(k_x) &= \tilde{h}_0\!\left(k_x + \frac{4\pi}{3}\right).
\end{align}
Consequently, the model $H_{\text{ch}}$ is mapped onto a two‑band model
\begin{equation}
\tilde{h}(\tilde{k}_x) = -t\cos \tilde{k}_x\, \sigma_0 + \sqrt{3}t\sin \tilde{k}_x\, \tilde{\sigma}_z + J \tilde{\sigma}_x, 
\label{eq:two_band_H}
\end{equation}
with \(\tilde{k}_x\in[-\pi/a,\pi/a]\). Since $\tilde{H}_{\rm SOC}$ passes unchanged through $U_{2,3}$, we finally obtain Eqs.~\eqref{eq4} and \eqref{eq5}: $H(\bm k)$ is mapped exactly onto an effective Rashba--Zeeman model. This is a highly anisotropic 2D Rashba model with a Zeeman field: the gauge field of the helix provides the ``Rashba'' term $\sqrt{3}t\sin\tilde{k}_x\tilde\sigma_z$, the SOC provides $2\lambda\sin k_y\tilde\sigma_y$, and the Zeeman term $J\tilde\sigma_x$ is the HZF---it originates from the helical magnetic order, not from any net magnetization. In magnetic materials one typically has $t>J>\lambda$, so the model is strongly anisotropic. We emphasize that the Rashba--Zeeman model invoked in Ref.~\cite{Yamada2025a,okumura2026} requires a net magnetization to break $\mathcal{T}$, in sharp contrast to our model. 
The six bands of $H(\bm k)$ are obtained from the two bands $E_\pm=\epsilon_0\pm|\bm d|$ by zone folding, as shown in Fig.~\ref{Fig1}(b); the degeneracy points at $k_x=0,\pm\pi$ are purely band-folding effects.

\begin{figure*}[t]
\includegraphics[width=7in]{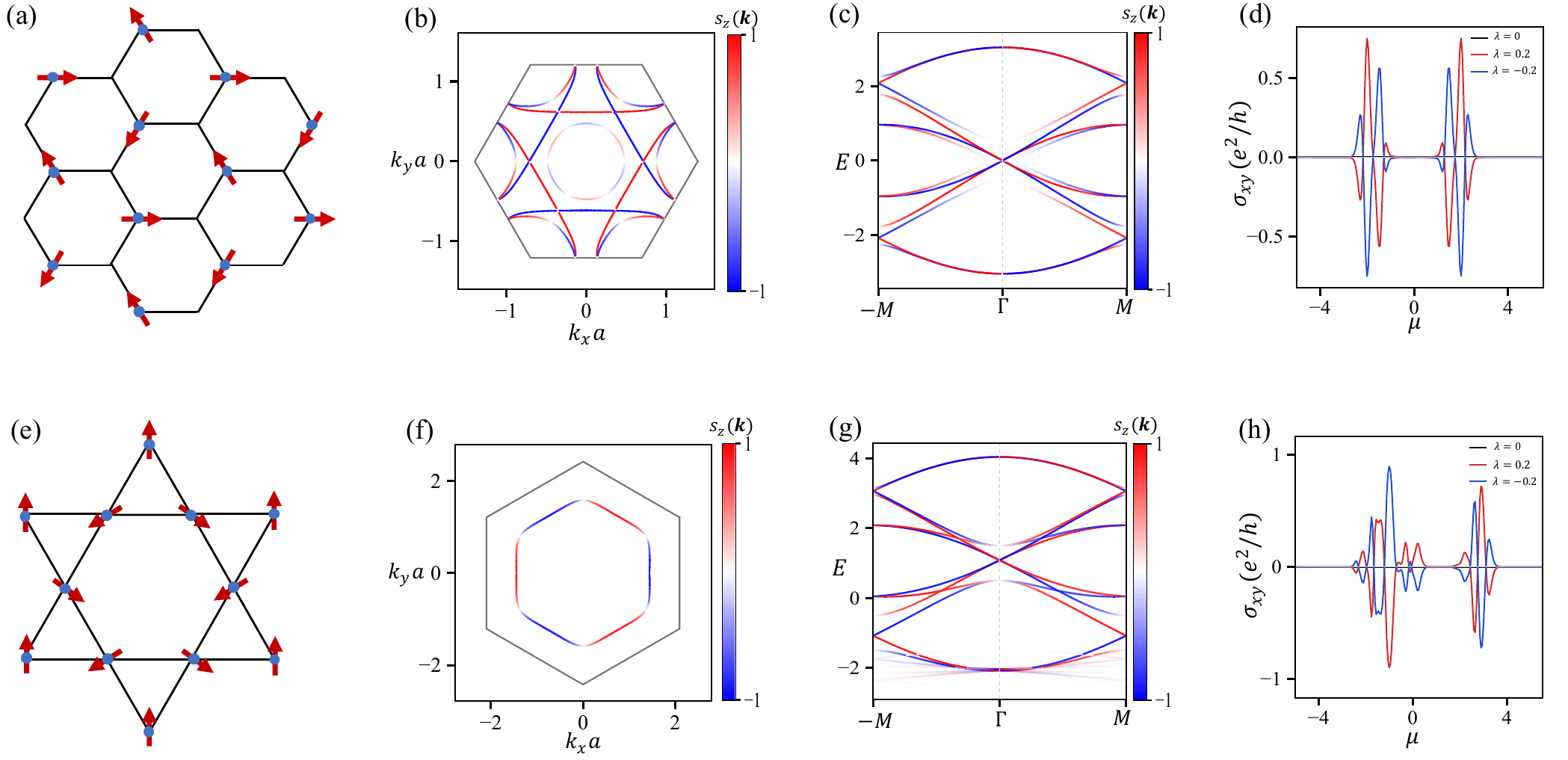}
\caption{FIG. 2. SOC-induced AHE in the honeycomb (a)--(d) and kagome (e)--(h) OPMs.
(a),(e) Real-space magnetic structure: a coplanar $120^\circ$ order with a
$\sqrt3\times\sqrt3$ magnetic cell. (b),(f) Fermi-surface contours at
$\lambda_R=0$, coloured by the out-of-plane spin expectation value $s_z(\bm k)$,
at $\mu=1.0$ (honeycomb) and $-0.5$ (kagome), inside the magnetic Brillouin
zone (hexagon). (c),(g) Band structures along $-M$--$\Gamma$--$M$ ($M$ a
time-reversal-invariant momentum) at $\lambda_R=0$, coloured by $s_z(\bm k)$.
(d),(h) Anomalous Hall conductivity $\sigma_{xy}(\mu)$ at $\lambda_R=0,\pm0.2$
computed with a small smearing $k_BT=0.05t$: the response vanishes identically
without SOC and reverses sign with $\lambda_R$. Model parameters are $t=1$ and
$J=0.5$.}
\label{Fig2}
\end{figure*}



\textit{AHE in helimagnets.}---Since the unitary transformations are momentum independent, the two-band model $\tilde{H}(\bm k)$ fully captures the AHE of ${H}(\bm k)$. The Berry curvature of the two-band model is
\beqn
\Omega_{\pm}(\bm{k}) &=& \pm\frac{1}{2} \hat{\bm{d}} \cdot (\partial_{\tilde{k}_x} \hat{\bm{d}} \times \partial_{k_y} \hat{\bm{d}}) \nonumber\\
&=& \pm\frac{\sqrt{3}\, t \lambda J \cos \tilde{k}_x \cos k_y}{\bigl( J^2 + 4\lambda^2 \sin^2 k_y + 3t^2 \sin^2 \tilde{k}_x \bigr)^{3/2}}.
\label{eq:berry_exact}
\eeqn
The curvature is peaked at the time-reversal-invariant momenta, where it reaches $|\Omega| = \sqrt{3}t\lambda/J^2$, and it vanishes identically for $\lambda=0$: SOC is the switch that converts the HZF-induced $\mathcal{T}$ breaking into Berry curvature. Figure~\ref{Fig1}(c) shows the distribution of $\Omega$ in the unfolded Brillouin zone. The anomalous Hall conductivity
\begin{equation}
\sigma_{xy} = \frac{e^2}{h} \sum_n\int_{\mathrm{BZ}} \frac{d^2\bm k}{2\pi} \; f\bigl(E_n(\bm{k})\bigr) \; \Omega_{n}(\bm{k})
\label{eq:sigma_xy_2D}
\end{equation}
is plotted in Fig.~\ref{Fig1}(d) as a function of the chemical potential $\mu$, for both the unfolded two-band model and the full six-band Hamiltonian; the two agree quantitatively, confirming the exactness of the mapping. 

Crucially, the AHE does not rely on the specific fine-tuned form of SOC. Figs.~\ref{Fig1}(e)--\ref{Fig1}(h) show the results for a uniform bond SOC $t_y=t\sigma_0+i\lambda\sigma_y$ on the interchain bonds, which no longer admits a simple analytic mapping: the Berry curvature and $\sigma_{xy}$ remain nonzero, with a reduced magnitude. The absence of a two-band mapping is evident from the band structure [Fig.~\ref{Fig1}(f)]: the band degeneracies are all lifted at the zone boundary, so the spectrum cannot be obtained by simple band folding. This confirms the generality of the mechanism---SOC breaks the $\{\mathcal{T}U_z(\pi)\,|\,I\}$ spin-group symmetry of the coplanar order, while the HZF retains the intrinsic $\mathcal{T}$ breaking.

\textit{SOC-induced AHE in honeycomb and kagome
OPMs.}---The same mechanism operates in
two lattice OPMs. (i) A coplanar $120^\circ$ order on the honeycomb lattice with
a $\sqrt{3}\times\sqrt{3}$ magnetic unit cell [Fig.~\ref{Fig2}(a)]. (ii) The
$\sqrt3\times\sqrt3$ $120^\circ$ coplanar
order on the kagome lattice, in which all nine sites of the
supercell carry a moment [Fig.~\ref{Fig2}(e)].  Both orders are
accessible in minimal correlated-electron models: the honeycomb order is the
mean-field state of the Hubbard model at filling one \cite{PhysRevB.109.L041103}, whereas the kagome order is the
$\sqrt3\times\sqrt3$ phase of the $J_1$--$J_2$ Heisenberg model \cite{fjjf-xspp}. We describe the magnetic phase
by the $sd$ Hamiltonian
\begin{equation}
H_{1,2}=t\!\!\sum_{\langle ij\rangle,\sigma}\!\!c_{i\sigma}^\dagger c_{j\sigma}
+J\sum_i\bm m_i\!\cdot\!c_i^\dagger\bm\sigma c_i,
\label{eq:sd_general}
\end{equation}
where the in-plane unit moments $\bm m_i=(\cos\theta_i,\sin\theta_i,0)$ are
parametrized by the angle $\theta_i$ of the local moment at site $i$, and $H_1$ ($H_2$) describes the honeycomb (kagome) order.  The
corresponding real-space magnetic structures $\bm m(\bm r)$ are shown in
Figs.~\ref{Fig2}(a) and \ref{Fig2}(e).

Two symmetries of these coplanar patterns fix the spin texture. (i) $H_1$ respects the magnetic symmetry
$C_3=\{U_z(2\pi/3)||R_3\}$ and $H_2$ the spin-group symmetry
$\mathcal{C}_3=\{I||R_3\}$; both give $s_z(R_3\bm k)=s_z(\bm k)$ and a vanishing in-plane spin expectation value, $s_{x,y}(\bm k)=0$. Here $R_n$ denotes
a rotation by $2\pi/n$ about the hexagon center. (ii) The spin-only
antiunitary operation $\{\mathcal{T}U_z(\pi)|I\}$ enforces $s_z(\bm k)=-s_z(-\bm k)$. Together
they enforce $s_z(R_6\bm k)=-s_z(\bm k)$, giving rise
to an $f$-wave magnet~\cite{2025luospin}. This texture is directly
visible in the Fermi-surface contours of Figs.~\ref{Fig2}(b) and \ref{Fig2}(f),
encoded by $s_z(\bm k)$: the polarization alternates in sign six times
around $\Gamma$ ($\bm k=0$) and reverses between $\bm k$ and $-\bm k$.

Notably, neither $H_1$ nor $H_2$ admits an analytic mapping of the kind derived above for the
helimagnet, yet their band structures share the same character as that of the
analytically solved helimagnet [Fig.~\ref{Fig1}], as
Figs.~\ref{Fig2}(c) and \ref{Fig2}(g) show: the coplanar order generates a
nonrelativistic spin splitting that is odd under momentum inversion,
$s_z(\bm k)=-s_z(-\bm k)$. However, the NSS alone, being compatible with an effective
time-reversal symmetry, cannot capture the intrinsic $\mathcal{T}$ breaking of the
order. The latter is reflected instead by the HZF, which is revealed by the lifting of the Kramers
degeneracy at the time-reversal-invariant momenta \cite{j9m4-386b}: at $M$, all bands of the nonmagnetic
band structure are twofold degenerate, whereas the coplanar magnetic order
splits two of the six pairs of the honeycomb model and three of the nine pairs of
the kagome model, the splitting being directly visible in
Figs.~\ref{Fig2}(c) and \ref{Fig2}(g). This
splitting, produced by the HZF without any net magnetization, captures the
intrinsic $\mathcal{T}$ breaking of the odd-parity order, and it is the essential ingredient
on which the AHE is built.

The AHE is switched on by the standard nearest-neighbor Rashba SOC
\begin{equation}
H_{\rm SOC}=i\lambda_R\!\!\sum_{\langle ij\rangle}\sum_{ss'}
c_{is}^\dagger\bigl[(\bm s\times\hat{\bm d}_{ij})_z\bigr]_{ss'}c_{js'},
\label{eq:SOC_rashba}
\end{equation}
where $\bm s$ is the spin operator, $\hat{\bm d}_{ij}$ the unit vector of the
bond $i\!\to\!j$, $\lambda_R$ the SOC strength, and the sum runs over
nearest-neighbour bonds. The Rashba term breaks
the residual $\{\mathcal{T}U_z(\pi)|I\}$ while preserving $C_3$, so no symmetry enforces zero Hall conductance. Figures.~\ref{Fig2}(d) and \ref{Fig2}(h) show $\sigma_{xy}(\mu)$ at fixed
$\lambda_R$ and we can see that $\sigma_{xy}$ is generally nonzero for $\lambda_R\neq 0$. We emphasize that both the HZF and SOC ingredients are
important---the HZF supplies the $\mathcal{T}$ breaking and the SOC supplies
the spin-space-group breaking---and neither alone is sufficient. 

\textit{Conclusion and discussion.}---In summary, we have established the symmetry condition and the microscopic
mechanism of the AHE in OPMs. The AHE originates from the interplay of two
ingredients: the HZF, rooted in the intrinsic $\mathcal{T}$ breaking of the
magnetic order, and SOC, which breaks the residual spin-group symmetry; neither
alone is sufficient. We have demonstrated this picture analytically through an
exact two-band mapping of a 2D helimagnet and numerically for OPMs on the
honeycomb and kagome lattices.

We note that the magnetic order reported for
Gd$_3$(Ru$_{1-\delta}$Rh$_{\delta}$)$_4$Al$_{12}$~\cite{Yamada2025a} preserves the
$\mathcal{T}\bm{\tau}$ symmetry. Since $\mathcal{T}\bm{\tau}$ is a symmetry of the
magnetic group, it cannot be broken by SOC, and within our framework it forbids
the AHE. The experimental observation of a giant AHE in this compound therefore
indicates that the true magnetic order cannot be a compensated helimagnet: a net
magnetization, or some other $\mathcal{T}\bm{\tau}$-breaking ingredient, must be
present.

\textit{Note added.}---After completing this work, we became aware of a related
preprint~\cite{bai2026timereversaloddtranspor} studying time-reversal-odd
transport in OPMs.

\begin{acknowledgments}
Xun-Jiang Luo thanks Xilin Feng and Ziting Sun for helpful
discussions. This work was supported by the National Key R\&D
Program of China (Grant Nos. 2024YFB3614100), the National
Natural Science Foundation of China (Grants Nos. 12604254,
12241405, 12274411), the Basic Research Program of the Chinese
Academy of Sciences Based on Major Scientific Infrastructures
(Grant No. JZHKYPT-2021-08), and the CAS Project for Young
Scientists in Basic Research (Grant No. YSBR-084).
\end{acknowledgments}

\appendix

\section{Appendix A: Candidate OPMs hosting the AHE}
\label{appendix}
Combining the spin-group symmetry criteria for OPMs~\cite{2025luospin,zy7s-j86r}
with the ferromagnetic-point-group condition established in the main text, we
identify the candidate OPMs listed in Table~\ref{tab:candidates}. These
materials are independently tabulated as spin-orbit magnets in the Supplementary
Material of Ref.~\cite{Yuntian2025}, whose columns give the material ID in
MAGNDATA, the chemical formula, the oriented spin space group (OSSG), the
magnetic space group (MSG), whether SOC induces a Wyckoff splitting, and the
MSG-allowed uniform magnetization $\bm M$. A nonzero allowed $\bm M$ is precisely
the ferromagnetic-point-group condition under which our mechanism permits a
field-free AHE. Most of the listed compounds are
triangular-lattice $120^\circ$ magnets or helimagnets of the coplanar type
discussed in the main text; EuIn$_2$As$_2$ and RbNiCl$_3$, both satisfying the
symmetry condition. 
\begin{table}[h]
\caption{Candidate OPMs whose magnetic point group is ferromagnetic and which
can therefore host a field-free AHE. The columns follow the spin-orbit-magnet
tabulation of Ref.~\cite{Yuntian2025}: MAGNDATA ID, magnetic space group (MSG)
with its BNS number, whether SOC induces a Wyckoff splitting, and the
MSG-allowed uniform magnetization $\bm M$ in the conventional basis of the spin
space group.}
\label{tab:candidates}
\begin{ruledtabular}
\begin{tabular}{llccc}
Material & ID & MSG & Wyckoff split & Allowed $\bm M$ \\
\hline
CsNiCl$_3$ & 1.0.42 & $C22'2_1'$ (20.34) & Yes & $(0,m_2,0)$ \\
RbNiCl$_3$ & 1.0.41 & $C22'2_1'$ (20.34) & Yes & $(0,m_2,0)$ \\
EuIn$_2$As$_2$ & 1.0.32 & $P6_12'2'$ (178.159) & Yes & $(0,0,m_3)$ \\
Ba$_3$CoSb$_2$O$_9$ & 1.0.44 & $Cm'c2_1'$ (36.174) & Yes & $(0,m_2,0)$ \\
CsMnBr$_3$ & 1.0.35 & $P\bar{6}2'm'$ (189.225) & No & $(0,0,m_3)$ \\
\end{tabular}
\end{ruledtabular}
\end{table}

\bibliography{reference,reference1}

\end{document}